\documentclass[pra,twocolumn,superscriptaddress,10pt,noshowpacs]{revtex4}
\usepackage[english]{babel}
\usepackage[T1]{fontenc}
\usepackage[utf8]{inputenc}
\usepackage{graphicx,epstopdf}
\usepackage{amssymb}
\usepackage{amsmath}
\usepackage{float}
\usepackage{bbm}

\usepackage{amsfonts}
\usepackage{bbm}
\usepackage{color}
\usepackage{latexsym}
\usepackage{caption}
\usepackage{subcaption}
\usepackage{times,txfonts}
\usepackage{color,soul}
\usepackage{listings}
\usepackage{bbold}
\usepackage{mathrsfs}
\usepackage[pdftex,plainpages=false,colorlinks=true,citecolor=blue,linkcolor=blue,urlcolor=blue,filecolor=green,bookmarksopen=true]{hyperref}
\usepackage{hyperref}

\begin{document}
	
	\title{Dirac fermions on a surface with localized strain}
	
	\author{Samuel B. B. Almeida}
	\email{samuelbastos@fisica.ufc.br}
	\affiliation{Universidade Federal do Cear\'a, Departamento de F\'{i}sica, 60455-760, Fortaleza, CE, Brazil}
	
	\author{J. E. G. Silva}
	\email{euclides@fisica.ufc.br}
	\affiliation{Universidade Federal do Cear\'a, Departamento de F\'{i}sica, 60455-760, Fortaleza, CE, Brazil}
	
	\author{C. A. S. Almeida}
	\email{carlos@fisica.ufc.br}
	\affiliation{Universidade Federal do Cear\'a, Departamento de F\'{i}sica, 60455-760, Fortaleza, CE, Brazil}

	\date{\today}
	
	\begin{abstract}
		We study the influence of a localized Gaussian deformation on massless Dirac fermions confined to a two-dimensional curved surface. Both in-plane and out-of-plane displacements are considered within the framework of elasticity theory. These deformations couple to the Dirac spinors via the spin connection and the vielbeins, leading to a position-dependent Fermi velocity and an effective geometric potential. We show that spin connection modifies the density of states near the origin and how this modification is altered by changing mechanical parameters. Analytical and numerical solutions reveal the emergence of asymptotically free states, but with changes in amplitude near the origin due to the modification of curvature mediated by Lamé coefficients. Upon introducing an external magnetic field, the effective potential becomes confining at large distances, producing localized Landau levels that concentrate near the deformation, thus enabling the evaluation of how mechanical coefficients affect the localization of states. A geometric Aharonov-Bohm phase is identified through the spinor holonomy. These results contribute to the understanding of strain-induced electronic effects in Dirac materials, such as graphene.
			\end{abstract}
	
	
	\maketitle
	\section{Introduction}
	Two-dimensional systems \cite{geim2009graphene,kara2012review,carvalho2016phosphorene} provide an exceptional environment for studying relativistic physics, with direct connections to gravitation and topology. The presence of curvature, spin-orbit couplings \cite{liang2018pseudo, shitade2020geometric, eremko2020generalized}, and anomalous effects makes these systems a rich environment for investigating phenomena that would traditionally be associated with fundamental particle and field theories. 
	These systems serve as gravitational analogs, providing insights into high-energy and topological phenomena in condensed matter contexts \cite{gallerati2022graphene, sepehri2016emergence}. In the context of fermions, curvature modifies the spin connection, thereby affecting the Dirac equation. 
	Since curvature alters the density of electronic states and the local Fermi velocity, this opens up an area of applications called straintronics \cite{si2016strain, sahalianov2019straintronics} where mechanical deformations rather than electric fields control the current. The most relevant case is certainly deformed graphene \cite{geim2007rise,katsnelson2,katsnelson,electronic1}, where the spin connection contributes terms that are effectively interpreted as gauge fields. Among the characteristics of graphene, the most relevant in the context of this work is that the electrons in the hexagonal carbon lattice behave effectively as relativistic particles. 
	 They can thus be modeled using the Dirac equation. In 2005, measurements were made of the anomalous quantum Hall effect \cite{novoselov2005two}, which indicated the presence of electrons behaving like relativistic particles with linear band structures. Due to these already well-determined properties, we can treat the particles as massless. With this in mind, we focused on the treatment of massless fermions.
	In principle, there are two main ways of analyzing these systems, the first consists of using solid-state physics methods, such as tight-binding \cite{tb1,tb2,tb3,tb4,ribeiro2009strained,manes2013generalized}, treating graphene as a crystalline lattice; the second is more closely associated with QFT methods in curved spaces \cite{birrell}, in general, one starts with the effective Dirac equation for graphene and, by considering a curved metric, one obtains the gauge fields induced by the geometry \cite{vozmediano1}. In particular, the latter was the first to predict a space-dependent Fermi velocity, which was confirmed experimentally \cite{vozmediano3}. In this context, the Dirac equation has been studied extensively for different geometries \cite{furtado, atanasov2015helicoidal, watanabe2015electronic, atanasov2010tuning, flouris2022curvature, carlos, yecsiltacs2018dirac}.
	In this work, we use the continuous approximation to deal with a smooth surface and then employ the differential approach \cite{BJ,diracsurface} to realize the adaptation of the Dirac equation in (2+1) dimensions, then we introduce deformations in the plane through the displacement vector, as suggested in \cite{arias1}. We extend the geometric approach by explicitly incorporating the effects of deformation through the theory of elasticity, with a particular focus on the role of Lamé coefficients in the dynamics of Dirac fermions. Considering a Gaussian deformation as in \cite{vozmediano1}, we check the field-curvature relationship and obtain an analytical expression for the pseudomagnetic field induced by the deformation in terms of strain and curvature vectors.  
	We also investigated the influence of an external magnetic field on the emergent potentials and the density of states. In both cases, we demonstrate the emergence of Aharonov-Bohm-like phases, which signal a distinction between the effective field and the true gauge field.
	This work is organized as follows: in section \ref{sect2}, we describe and analyze the geometry of the problem, introducing the displacement vectors, and we also examine the relationship between curvature and the effective gauge field produced by the deformations. In section \ref{sect4}, we perform an adaptation of the Dirac equation to obtain effective potentials, where in \ref{sec3} we obtained numerically stationary solutions and some analytical approximations. Finally, in section \ref{sect5} we introduce an external magnetic field to observe how it alters the potentials and densities of states obtained in \ref{sect4}. In appendix \ref{relationworks}, we presented a way of calculating some constants from the purely geometric approach in terms of Lamé constants. Some results that are too long and not very practical can be found in the appendix \ref{apendd}.
	
	\section{The geometry of the gaussian bump}\label{sect2}
	
	We investigate the combined effects of in-plane and out-of-plane deformations in a graphene sheet using a continuous approximation, i.e., modeling the sheet as a smooth surface. We will consider an out-of-plane deformation in the form of a Gaussian
	\begin{equation}\label{01}
		h(r)=h_0e^{-\frac{r^2}{b^2}}\,,
	\end{equation}
	where $r=\left(x^2+y^2\right)^{1/2}$. Taking the gradient of equation \eqref{01}, it is possible to employ the approach of reference \cite{vozmediano1} and introduce the perturbation directly into the metric. However, in this work, the function $\partial_rh(r)$ will be taken as only one of the deformation contributions. Since, from a mechanical point of view, it induces an in-plane deformation field $u_r$. Both types of deformation will be introduced into the metric via elasticity theory, namely with the strain tensor
	\begin{equation}\label{02}
		u_{\mu\nu}=\frac{1}{2}(\partial_{\mu}u_{\nu}+\partial_{\nu}u_{\mu}+2h_{\mu}h_{\nu})\,,
	\end{equation}
	$u_\mu$ and $u_\nu$ denote the intrinsic displacements and $h_{\mu}$ and $h_{\nu}$ are the extrinsic ones, given by
	\begin{equation}
		h_\nu=\partial_{\nu}h\,,
	\end{equation}
	and since the problem is defined on a surface, all the indices up to now have only two values. It has been show that the most suitable coordinates are cylindrical, we adopt cylindrical coordinates $\mu=r,\theta$. 
	According to the elasticity theory \cite{landau}, the metric is related to the strain tensor as follows
	\begin{equation}
			ds^2=g_{\mu\nu}dx^\mu dx^\nu\,, \text{with}
	\end{equation}
	\begin{equation}\label{03}
		g_{\mu\nu}=\delta_{\mu\nu}+2u_{\mu\nu}\,.
	\end{equation}
	we use the following signature Minkowski $\operatorname{sign}(\eta) =(+--)$.
	
	Based on the results of the reference \cite{arias1}, we will propose that the displacements in the plane associated with the Gaussian deformation are of the form
	\begin{equation}\label{04}
		\scalebox{0.85}{$\displaystyle \begin{aligned}
				u_r=-\frac{\sqrt{\frac{\pi }{2}} h_0^2 r e^{-\frac{2 r^2}{b^2}} \left(b^2 \left(e^{\frac{2 r^2}{b^2}}-1\right) (\lambda +\mu )-2 r^2 (\lambda +2 \mu )\right)}{b^2 r^2 (\lambda +2 \mu )}\,,
			\end{aligned}$}
	\end{equation}
	the terms $\mu$ and $\lambda$ are the Lamé coefficients.
	\begin{figure}[H]
		\centering
		\includegraphics[scale=0.72]{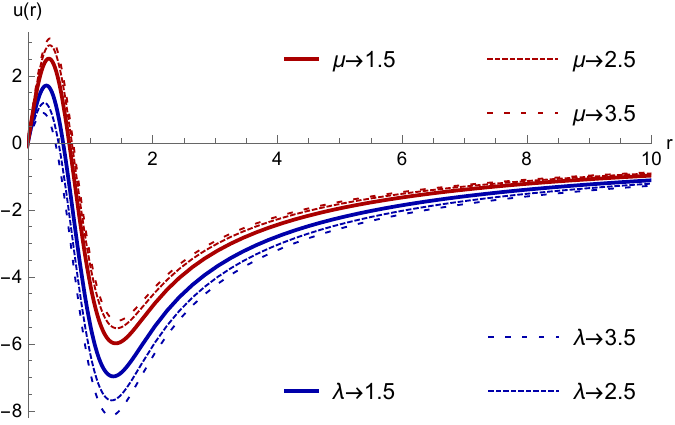}
		\caption{Behavior of variations in $\lambda$ and $\mu$, given $h_0$=5, $b$=1, $\mu$=1.}
		\label{fig07}
	\end{figure}
	
	The coefficients $\mu$ and $\lambda$ behave differently; more specifically, they tend to have opposite contributions, since as $\mu$ increases, the $u(r)>0$ contributions tend to increase, while increases in $\lambda$ tend to increase the $u(r)<0$ contributions. 
	In cylindrical coordinates we have 
	\begin{equation}
		ds^2=(1+2u_{rr})dr^2+(1+2u_{\theta\theta})r^2d\theta^2+4u_{r\theta}rdrd\theta\,,
	\end{equation}
	 and, since we only have the radial contribution, the strain tensor \eqref{02} can be written in cylindrical coordinates \cite{landau}, for $u_\theta=0$, as 
	\begin{equation}
		\begin{aligned}
			&u_{rr}=\partial_r u_r+(\partial_rh)^2\\
			&u_{\theta\theta}=\frac{u_r}{r}\\
			&u_{r\theta}=0\,.
		\end{aligned}
	\end{equation}
	Using these equations, we will have
	\begin{equation}\label{004}
		\scalebox{0.75}{$\displaystyle  \begin{aligned}
				u_{rr}=\frac{h_0^2 e^{-\frac{2 r^2}{b^2}} \left(-2 \sqrt{2 \pi } b^2 \lambda  r^2+\sqrt{2 \pi } b^4 \left(e^{\frac{2 r^2}{b^2}}-1\right) (\lambda +\mu )-8 \left(\sqrt{2 \pi }-2\right) r^4 (\lambda +2 \mu )\right)}{4 b^4 r^2 (\lambda +2 \mu )}
			\end{aligned}$}
	\end{equation}
	and
	\begin{equation}
		u_{\theta\theta}=-\frac{\sqrt{\frac{\pi }{2}} h_0^2 e^{-\frac{2 r^2}{b^2}} \left(b^2 \left(e^{\frac{2 r^2}{b^2}}-1\right) (\lambda +\mu )-2 r^2 (\lambda +2 \mu )\right)}{2 b^2 r^2 (\lambda +2 \mu )}\,.
	\end{equation}
	So the metric, including the time component, is $g_{\mu\nu}=\operatorname{diag}{(1,-g_{rr},-g_{\theta\theta})}$ 
	with components
	\begin{equation}\label{07}
		\scalebox{1}{$\displaystyle \begin{aligned}
				g_{rr}=1+\frac{h_0^2 e^{-\frac{2 r^2}{b^2}} \left(-2 \sqrt{2 \pi } b^2 \lambda  r^2+\sqrt{2 \pi } b^4 \left(e^{\frac{2 r^2}{b^2}}-1\right) (\lambda +\mu )\right)}{2 b^4 r^2 (\lambda +2 \mu )}-\\
				-h_0^2 e^{-\frac{2 r^2}{b^2}} \frac{8 \left(\sqrt{2 \pi }-2\right) r^4 (\lambda +2 \mu )}{2 b^4 r^2 (\lambda +2 \mu )}
			\end{aligned}$}
	\end{equation}
	\begin{equation}\label{08}
		g_{\theta\theta}=r^2+\frac{\sqrt{\frac{\pi }{2}} h_0^2 e^{-\frac{2 r^2}{b^2}} \left(2 r^2 (\lambda +2 \mu )-b^2 \left(e^{\frac{2 r^2}{b^2}}-1\right) (\lambda +\mu )\right)}{b^2 (\lambda +2 \mu )}\,.
	\end{equation} 
	\begin{figure}[H]
		\centering
		\includegraphics[scale=0.07]{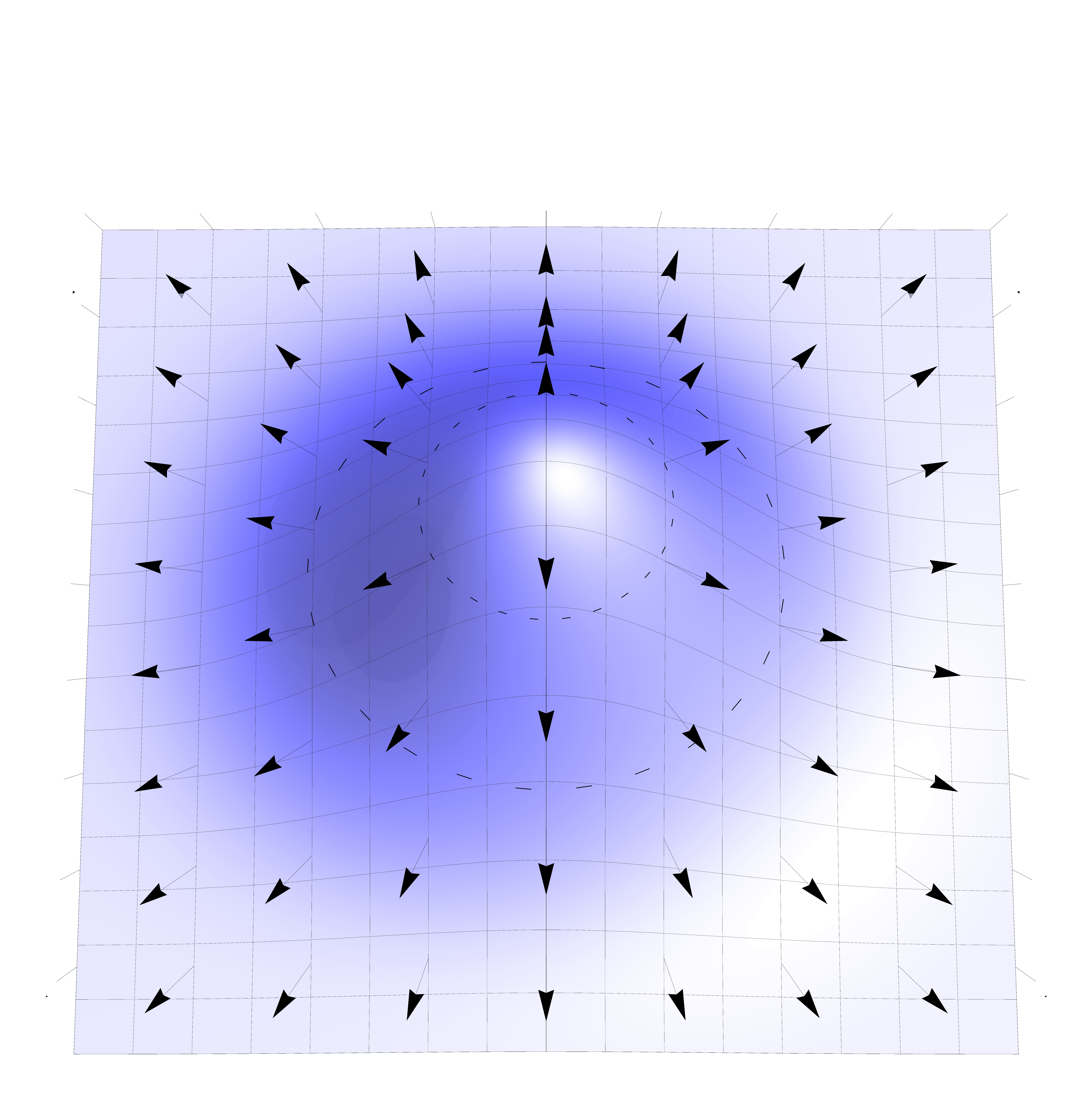}
		\caption{Field of vectors defined from $u_r$ on the Gaussian surface.}
		\label{fig37}
	\end{figure}
	The change in metrics due to elastic parameters can be seen in figures \ref{fig10} and \ref{fig11}.  It is interesting to note that, in addition to the radial component, the angular component of the metric is also modified, which is not particularly expected since the geometric shape has angular symmetry. This change is due to the nature of the strain tensor, which, in cylindrical coordinates \cite{landau,soutas2012elasticity}, includes changes in $g_{\theta\theta}$ even though the displacement vector is only radial.
	\begin{figure}[H]
		\centering
		\includegraphics[scale=0.67]{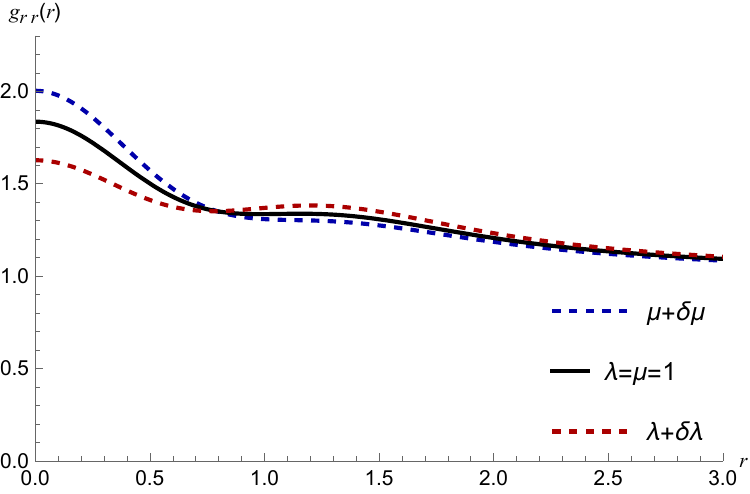}
		\caption{Variations of the radial component of the metric with $\lambda$ and $\mu$. Where $\delta\lambda=\delta\mu=1$. Here $h_0=b=1$.}
		\label{fig10}
	\end{figure}
	
	\begin{figure}[H]
		\centering
		\includegraphics[scale=0.67]{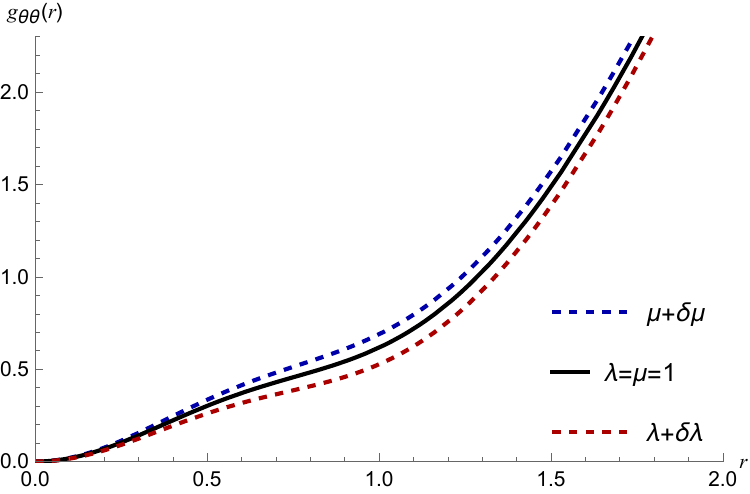}
		\caption{Variations of the angular component of the metric with $\lambda$ and $\mu$. Where $\delta\lambda=\delta\mu=1$. Here $h_0=b=1$.}
		\label{fig11}
	\end{figure}
	We also derive approximate expressions valid near the origin (small r). In this case, we have
	\begin{equation}\label{28}
		g_{rr}\approx1+C_1r^2+C_2
	\end{equation}
	and
	\begin{equation}\label{29}
		g_{\theta\theta}\approx C_3r^2\,,
	\end{equation}
	where the constants are given by \eqref{ccst01}, \eqref{ccst02} and \eqref{ccst03}. Here, it is worth noting that we can define the following term in \eqref{004} $\alpha=\frac{h_0}{b}$. So we will have
	\begin{equation}
		g_{rr}\approx1+\alpha^2r^2C'_1+\alpha^2C'_2\coloneqq\mathfrak{g}_{rr}\,,
	\end{equation}
	in which $C'_1$ and $C'_2$ are the constants redefined by factoring $\alpha^2$. Rewriting it this way, just as we did in \cite{vozmediano1}, we can look at case $b>>h_0$. So we immediately see that
	\begin{equation}\label{27}
		\lim_{\alpha\to 0} \mathfrak{g}_{rr}=\lim_{h_0\to0}\mathfrak{g}_{rr}=1\,.
	\end{equation}
	This will be useful for determining analytical approximations in section \ref{sec3}.
	
	We will use Greek indices for the curved space and initial Latin indices to denote the apartment space. Let's consider that there is local Lorentz symmetry, so we can relate the curved metric $g_{\mu\nu}$ to the flat Minkowski metric $\eta_{ab}$ $(+--)$
	\begin{equation}
		\label{metricvielbein}
		g_{\mu\nu}=e^a_\mu e^b_\nu\eta_{ab}\,,
	\end{equation}
	where the vielbeins satisfy $e^b_{\mu}e_a^{\mu}=\delta^b_{a}$.
	The flat indices will take the values $a=0,1,2$. A general choice for the vielbein $e^a_\mu $ satisfying Eq. (\ref{metricvielbein}) has the form
	\begin{equation}\label{06}
		e^a_{\mu} =
		\begin{pmatrix}
			1 & 0 & 0 \\
			0 & \sqrt{g_{rr}}\cos{\theta} & -\sqrt{g_{\theta\theta}}\sin{\theta} \\
			0 & \sqrt{g_{rr}}\sin{\theta} & \sqrt{g_{\theta\theta}}\cos{\theta}
		\end{pmatrix}\,,
	\end{equation}
	
	Then we can obtain the connection 1-form $\omega^c_a=\omega^c_{\mu a}dx^\mu$ by
	\begin{equation}
		\omega^{ab}_\mu=e^a_\mu\Gamma^\nu_{\sigma\mu}+e^a_\nu\partial_\mu e^{\nu b}\,,
	\end{equation}
	where the Christoffel symbols are defined as
	\begin{equation}
		\Gamma^\rho_{\mu\nu} = \frac{1}{2} g^{\rho\sigma} 
		\left( \partial_\mu g_{\sigma\nu} 
		+ \partial_\nu g_{\sigma\mu} 
		- \partial_\sigma g_{\mu\nu} \right)\,.
	\end{equation}
	
\begin{figure}[H]
		\centering
		\includegraphics[scale=0.85]{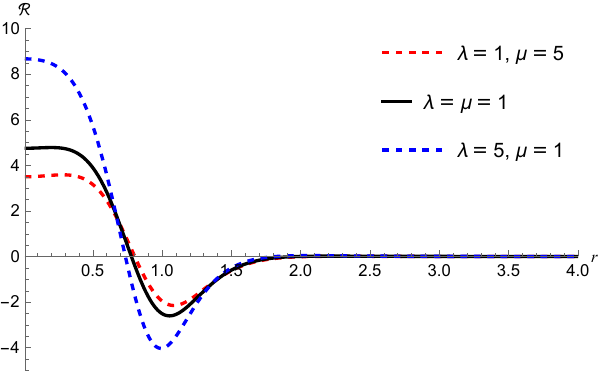}
		\caption{Variations $\delta\lambda=\delta\mu=1$. For the continuous curve $h_0=1$, $b=1$, $\lambda=\mu=1$.}
		\label{fig09}
	\end{figure}

	The results obtained are lengthy, and they can be found in the appendix. Namely, if we do not take into account the contributions of the displacement vectors $u_r$, the Christoffel symbols \eqref{crsft01}, \eqref{crsft02} and \eqref{crsft03} are reduced to those obtained in \cite{vozmediano1}, except for the symmetrization factor $2$ in \eqref{03}, i.e., for $u_r\to0$ we have  
	\begin{equation}
		\scalebox{1}{$\displaystyle \begin{aligned}
				&\Gamma^\theta_{r\theta}=\frac{1}{r}\\
				&\Gamma^r_{\theta\theta}=-\frac{r}{1+\gamma f(r)}\\
				&\Gamma^r_{rr}=\frac{1}{2}\frac{\gamma \partial_rf(r)}{(1+\gamma f(r))}\,,
			\end{aligned}$}
	\end{equation}
	where here $\gamma=\frac{8h^2_0}{b^4}$ and $f(r)=r^2e^{-\frac{2r^2}{b^2}}$.
	For our static case and with $u_\theta=0$ and $\partial_\theta h=0$, the curvature can be written in terms of the deformation vector by \ref{24}. The analytical expression for $\mathcal{R}(r)$ is too long and is not very useful. However, it is possible to numerically analyze, in figure \ref{fig09}, the behavior due to variations in Lamé coefficients. 
	
	Near the origin, it is important to note again the opposite contributions of $\lambda$ and $\mu$ to the curvature. Now we can already adapt the Dirac equation. In particular, we will be interested in the search for stationary states, so we will now determine the Hamiltonian adapted to the geometry of the problem.
	
	The relationship between effective fields and curvature is a well-known fact in the literature \cite{vozmediano1,vozmediano2}. Following the approach used in \cite{arias1}, we identify the non-zero component of the spin connection by the potential geometric vector $\mathcal{S_\theta}=\omega^{12}_\theta$ vector such that
	\begin{equation}\label{25}
		\mathcal{B}=\frac{1}{\sqrt{|g|}}\varepsilon^{\mu\nu}\nabla_\mu \mathcal{S_\nu}\,.
	\end{equation}
	
	For different variations of $\lambda$ and $\mu$ the relationship $2\mathcal{B}=a\mathcal{R}$ is satisfied for $a=1$. In other words, since in two dimensions the scalar curvature is half the Gaussian curvature $\mathcal{K}(r)$, we have that
	\begin{equation}
		\mathcal{B}=\mathcal{K}\,,
	\end{equation}
	and this result coincides with \eqref{25} applied to $\omega^{12}_\theta$ and $\mathcal{K}$ obtained in \cite{euclides1}. In work \cite{arias1}, Arias obtains the relation $\mathcal{B}=\mathcal{R}$, but bases it on the spin connection calculated in \cite{vozmediano1}, which omits the factor $2$ of symmetrization in the product of the Dirac matrices. Absorbing this factor reduces $\mathcal{B}=\mathcal{R}/2$ to $\mathcal{B}=\mathcal{R}$. But here we use the fact that in two dimensions the scalar curvature is equal to twice the Gaussian curvature.
	
	\section{Fermion dynamics}\label{sect4}
	Assuming that the spinor is contained in a space of (2+1) dimensions and that, like many similar models \cite{vozmediano1,arias1}, it has no mass and so will be treated as Weyl fermions with two components \cite{alisultanov2023non, nissinen2017type} , we obtain
	\begin{equation}\label{10}
		i\hbar\gamma^{\mu}(\partial_{\mu}+\Omega_{\mu})\Psi=0\,.
	\end{equation}
	
\begin{figure}[H]
		\centering
		\includegraphics[scale=0.8]{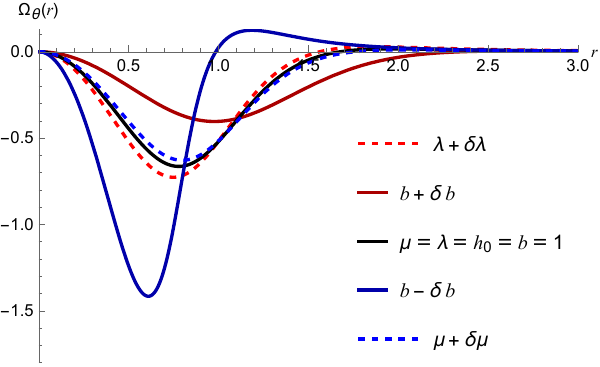}
		\caption{Spin connection, fixed $h_0=1$. Here $\delta\lambda=\delta\mu=2$ and $\delta b=0.3$ .}
		\label{fig14}
	\end{figure}

	We introduce curvature into the spinor derivative through the covariant derivative $\nabla_\mu=\partial_\mu+\Omega_\mu$. The following representation was adopted for the Dirac matrices $\gamma_0=\sigma_3$, $\gamma_1=-i\sigma_2$, and $\gamma_2= i\sigma_1$. As this is an effective approach, the light speed is replaced by the Fermi speed $v_F$. Expanding the sum by \eqref{10} gives us
	\begin{equation}
		i\hbar\frac{\partial \Psi}{\partial t}=-i\hbar v_F\gamma_0[\gamma^{r}\partial_r+\gamma^\theta(\partial_\theta+\Omega_\theta)]\Psi\,,
	\end{equation}
	where we identify the Dirac Hamiltonian by
	\begin{equation}\label{11}
		\mathcal{H}=-i\hbar v_F\sigma_3[\gamma^{r}\partial_r+\gamma^\theta(\partial_\theta+\Omega_\theta)]\,.
	\end{equation}
	The $\Omega_r$ component is null, so we only have contributions from $\Omega_\theta$. We have
	\begin{equation}\label{12}
		{\Omega}_\theta=\frac{1}{4}\omega^{21}_\theta\gamma_2\gamma_1\,,
	\end{equation}
	and we can analyze the behavior of $\Omega_\theta$ in figure \ref{fig14}.
	
	The curved Dirac matrices can be determined through contractions with vielbeins
	\begin{equation}
		\gamma^\mu=e^\mu_a\gamma^a\,.
	\end{equation}
	So we obtain the following expressions
	\begin{equation}
		\begin{aligned}
			&\gamma^t=\gamma^0\\
			&\gamma^r=\frac{\cos{\theta}}{\sqrt{g_{rr}}}\gamma^1+\frac{\sin{\theta}}{\sqrt{g_{rr}}}\gamma^2\\
			&\gamma^\theta=-\frac{\sin{\theta}}{\sqrt{g_{\theta\theta}}}\gamma^1+\frac{\cos{\theta}}{\sqrt{g_{\theta\theta}}}\gamma^2\,.
		\end{aligned}
	\end{equation}
	Using these matrices and the Hamiltonian \eqref{12} we get
	\begin{equation}
		\scalebox{0.74}{$\displaystyle \begin{aligned}
				\mathcal{H}=-i\hbar v_F\bigg[ \left( \frac{\cos{\theta}}{\sqrt{g_{rr}}}\sigma_1-\frac{\sin{\theta}}{\sqrt{grr}}\sigma_2\right)\partial_r-\left(\frac{\sin{\theta}}{\sqrt{g_{\theta\theta}}}\sigma_1+\frac{\cos\theta}{\sqrt{g_{\theta\theta}}}\sigma_2\right)\partial_\theta +\\
				+\frac{1}4\left(\frac{\sin{\theta}}{\sqrt{g_{\theta\theta}}}\sigma_2-\frac{\cos{\theta}}{\sqrt{g_{\theta\theta}}}\sigma_1\right)\omega^{21}_\theta \bigg] \,.
			\end{aligned}$}
	\end{equation}
	In particular, we will fix $\theta=0$, which ensures that the $\Gamma_\theta$ contributions go to zero for $h_0\to0$. Then
	\begin{equation}\label{14}
		\mathcal{H} =
		-i\hbar v_F\begin{pmatrix}
			0 & \frac{\partial_r}{\sqrt{g_{rr}}}+i\frac{\partial_\theta}{\sqrt{g_{\theta\theta}}}+\Gamma_\theta \\
			\frac{\partial_r}{\sqrt{g_{rr}}}-i\frac{\partial_\theta}{\sqrt{g_{\theta\theta}}}+\Gamma_\theta & 0
		\end{pmatrix}\,.
	\end{equation}
	where we are labeling the term that arises from the spin connection
	by
	\begin{equation}\label{30}
		\Gamma_\theta(r)=\frac{1}{4}\frac{\omega^{12}_\theta}{\sqrt{g_{\theta\theta}}}\,.
	\end{equation}
	Note that $\Gamma_\theta(r)$ acts as a geometric potential in the Hamiltonian. Looking at the expression \eqref{12}, we can conclude that, for $h_0=0$, this term vanishes. This is only true because once $\Gamma_\theta$ arises from the spin connection, this boundary condition is completely sensitive to the choice of vielbeins. For example, if \eqref{06} is explicitly of the form $\operatorname{diag}{(1,\sqrt{g_{rr}},\sqrt{g_{\theta\theta}})}$, then, even if we fix $h_0=0$, there will still be contributions that decrease with $r$, namely $\lim_{h_0 \to 0}\Gamma_\theta\propto \frac{1}{r}$. Since the vielbeins define local frames, these contributions are the effects of changes in a non-coordinate basis, so that the effective fields sense these choices. So, the choice $\theta=0$ guarantees that in the flat case, i.e. $h_0=0$, there is no geometric potential and the particle is completely free. Given that we have determined the Hamiltonian, the most natural way to study the effects of both deformations is to look for stationary states \cite{euclides1,euclides2,carlos}.
	
	\begin{figure}[H]
		\centering
		\includegraphics[scale=0.85]{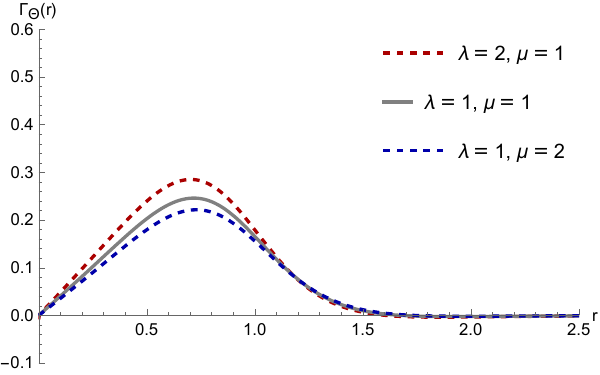}
		\caption{The geometric potential $\Gamma_\theta(r)$, fixed $b=1$. The curves in black, blue, and red show $h_0=1$. Here, $\delta\lambda=\delta\mu=1$.}
		\label{fig15}
	\end{figure}
	We can see that, for the above parameters, the potential is attractive near the origin but then becomes repulsive. Moreover, it has a finite barrier displayed and it vanishes asymptotically. For very small values of $b$, the potential diverges.
	
	\subsection{Stationary solutions} \label{sec3}
	
\begin{figure}[H]
		\centering
		\includegraphics[scale=0.845]{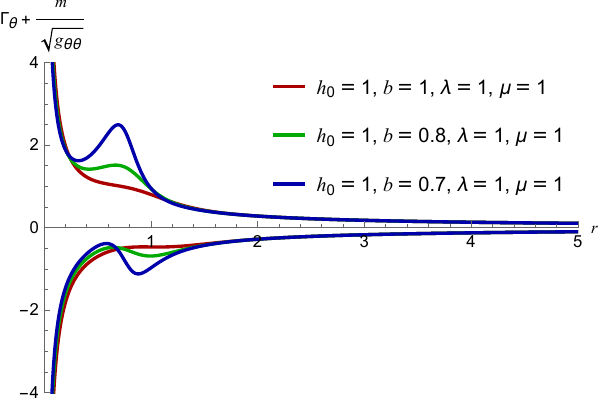}
		\caption{Behavior of $\Gamma_\theta(r)+\frac{m}{\sqrt{g_{\theta\theta}}}$. On the positive side, we have $m=\frac{1}{2}$ and in the negative, $m=-\frac{1}{2}$.}
		\label{figEuc01}
	\end{figure}

	We consider a separable solution of the form
	\begin{equation}
		\Psi(r,\theta,t)=e^{i\frac{E}{\hbar}t}e^{im\theta}\psi(r)
	\end{equation}
	\begin{equation*}
		\psi=\begin{pmatrix}
			\psi_1 \\ 
			\psi_2
		\end{pmatrix}\,,
	\end{equation*}
	such that
	\begin{equation}\label{15}
		\mathcal{H}\Psi=E\Psi\,.
	\end{equation}
	Developing this for the Hamiltonian \eqref{14}, we obtain two coupled equations
	\begin{equation}
		\begin{cases}
			\frac{1}{\sqrt{g_{rr}}}\partial_r\psi_2-\frac{m}{\sqrt{g_{\theta\theta}}}\psi_2+\Gamma_\theta\psi_2=\epsilon\psi_1\\
			\frac{1}{\sqrt{g_{rr}}}\partial_r\psi_1+\frac{m}{\sqrt{g_{\theta\theta}}}\psi_1+\Gamma_\theta\psi_1=\epsilon\psi_2
		\end{cases}\,,
	\end{equation}
	in which $\epsilon=\frac{E}{\hbar v_F }$. To decouple the equations, we can define the operators
	\begin{equation}
		\mathcal{O}_{1,2}=\frac{1}{\sqrt{g_{rr}}}\partial_r\pm\frac{m}{\sqrt{g_{\theta\theta}}}+\Gamma_\theta(r)\,.
	\end{equation}
	So we can interpret the $\frac{m}{\sqrt{g_{\theta\theta}}}+\Gamma_\theta$ term as an effective potential and then we notice that, from this perspective, the factor $\frac{1}{\sqrt{g_{\theta\theta}}}$ tends to zero near the origin, causing the behavior to become divergent. Namely, the behavior is governed by the geometric terms and by the parity of $m$, as shown in Figure \ref{figEuc01}.
	The factor $\sqrt{g_{\theta\theta}}$ .
	Thus \eqref{15} can be written in compact form by
	\begin{equation}
		\begin{pmatrix}
			0 & -i\mathcal{O}_2 \\
			-i\mathcal{O}_1 & 0
		\end{pmatrix} \begin{pmatrix}
			\psi_1 \\ 
			\psi_2
		\end{pmatrix}=\epsilon\begin{pmatrix}
			\psi_1 \\ 
			\psi_2
		\end{pmatrix}\,,
	\end{equation}
	such that
	\begin{equation}
		\begin{cases}
			-i\mathcal{O}_2\psi_2=\epsilon\psi_1\\
			-i\mathcal{O}_1\psi_1=\epsilon\psi_2
		\end{cases}\,.
	\end{equation}
	By multiplying the first by $-iO_1$ and the second by $-iO_2$, we can decouple the equations
	\begin{equation}
		\begin{aligned}
			&-\mathcal{O}_1\mathcal{O}_2\psi_2=\epsilon^2\psi_2\\
			&-\mathcal{O}_2\mathcal{O}_1\psi_1=\epsilon^2\psi_1
		\end{aligned}\, \, \, \, \,.
	\end{equation}
	Developing the first equation, we get
	\begin{equation}\label{16}
		-\frac{\partial^2_r\psi_2}{g_{rr}}+\left(-\frac{2\Gamma_\theta}{\sqrt{g_{rr}}}+\frac{1}{2}\frac{\partial_rg_{rr}}{g_{rr}^2}\right)\partial_r\psi_2+\overline{U}^2_2\psi_2=\epsilon^2\psi_2\,,
	\end{equation}
	with
	\begin{equation}
		\overline{U}^2_2=\left(\frac{m^2}{g_{\theta\theta}}-\frac{m\partial_rg_{\theta\theta}}{2g^{3/2}_{\theta\theta}\sqrt{g_{rr}}}-\frac{\partial_r\Gamma_\theta}{\sqrt{g_{rr}}}-\Gamma_\theta^2\right)\,.
	\end{equation}
	Similarly, the equation for $\psi_1$ is obtained by replacing $m\rightarrow-m$. In particular, we can define
	\begin{equation}\label{32}
		U^2_2=g_{rr}\overline{U}^2_2
	\end{equation}
	and
	\begin{equation}\label{17}
		\mathscr{V}_F(r)=\frac{v_F}{\sqrt{g_{rr}}}\,.
	\end{equation}
	We can call this the effective Fermi velocity; this coincides with the one defined in \cite{vozmediano1}. We have also
	\begin{equation}
		\varepsilon=\frac{E}{\hbar \mathscr{V}_F}=\frac{E\sqrt{g_{rr}}}{\hbar v_F}\,,
	\end{equation}
	then $\varepsilon(r)=\sqrt{g_{rr}}\epsilon$. Thus, multiplying \eqref{17} by $g_{rr}$ we have
	\begin{equation}\label{18}
		-\partial^2_r\psi_2+\left(-2\Gamma_\theta\sqrt{g_{rr}}+\frac{1}{2}\frac{\partial_rg_{rr}}{g_{rr}}\right)\partial_r\psi_2+U^2_2\psi_2=\varepsilon^2\psi_2\,.
	\end{equation}
	
	We employ the change on the wave function performed in reference \cite{euclides1} of the form $\psi_2(r)=\zeta(r)\chi_2(r)$, where $\zeta(r)$ is given by
	\begin{equation}\label{20}
		\zeta(r)=\zeta_0 e^{-\int \left( \Gamma_\theta\sqrt{g_{rr}} -\frac{1}{4}\frac{\partial_rg_{rr}}{g_{rr}}\right)dr}=\zeta_0e^{-\int \Gamma_\theta \sqrt{g_{rr}}dr}e^{\frac{1}{4}\ln{|g_{rr}|}}\,,
	\end{equation}
	in which $\zeta_0$ is a constant. In this way, we can obtain a Klein-Gordon-type expression with an associated squared effective potential. As a result, the new wave function $\chi_2 (r)$ satisfies the equation
	\begin{equation}\label{23}
		-\chi''_2+V^2_2\chi_2=\varepsilon^2\chi_2\,,
	\end{equation}
	
	and $V_{2}^{2}(r)$ is the effective squared potential and of form
	\begin{equation}\label{33}
		V^2_2=\frac{5}{16}\frac{(\partial_rg_{rr})^2}{g^2_{rr}}-\frac{1}{4}\frac{\partial^2_rg_{rr}}{g_{rr}}+m^2\frac{g_{rr}}{g_{\theta\theta}}-m\frac{\sqrt{g_{rr}}}{2}\frac{\partial_rg_{\theta\theta}}{g_{\theta\theta}^{3/2}}\,.
	\end{equation}
	
	It is worth noting the importance of the function $\zeta(r)$. The expression in Eq.(\ref{20}) depends on an integral of the geometric connection. Therefore, $\zeta$ can be understood as a kind of geometric Aharonov-Bohm effect, as discussed in \cite{furtado,carlos,geometricphase,cone}. In order to fix the constant $\zeta_0$, we consider $\lim_{h_0\to0}\zeta(r)=1$.
	
	Now let us analyze the behavior of the squared effective potential $V_{2}^{2}(r)$ and the respective wave functions. First, consider $V_{2}^{2}(r)$ far from the bump, whose expression becomes
	\begin{equation}
		V^2_2(r)\approx\frac{m^2}{r^2}-\frac{m r}{\left(r^2\right)^{3/2}}\,.
	\end{equation}
	Accordingly, the Klein-Gordon-like Eq.\eqref{23} yields
	\begin{equation}
		-\chi_2 ''(r)+\left(\frac{m^2}{r^2}-\frac{m r}{\left(r^2\right)^{3/2}}\right) \chi_2 (r)=\epsilon ^2 \chi_2 (r)\,,
	\end{equation}
	whose solution is given by
	\begin{equation}
		\scalebox{1}{$\displaystyle \begin{aligned}
				\chi_2(r) = \sqrt{r} \, A_1 \, J_{\frac{1}{2}(-1 + 2m)}(r \varepsilon) + \sqrt{r} \, A_2 \, Y_{\frac{1}{2}(-1 + 2m)}(r \varepsilon)\,,
			\end{aligned}$}
	\end{equation}
	where $J_n$ are Bessel functions of the first kind and $Y_n$ are the second kind. To avoid a divergence at the origin, we set $A_2 =0$. Therefore, the wave function behaves as a free state asymptotically, as expected.
	\begin{figure}[H]
		\centering
		\includegraphics[scale=0.83]{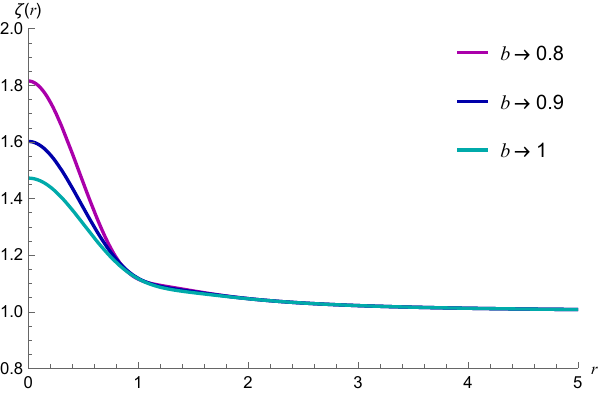}
		\caption{As expected, $\zeta(r)$ converges to 1 and has more influence near the origin. This result also shows that for localized deformations we should not expect localized states, since asymptotically we should have free particle solutions. Here $h_0=\lambda=\mu=1$ and $b$ is varying, showing that the lower its value, the more pronounced the phase is.}
		\label{fases}
	\end{figure}
	We normalize solutions with
	\begin{equation}
		|A|^2\int \sqrt{\operatorname{det} g} \ |\psi(r)|^2 rdr=1\,.
	\end{equation}
	where the angular contribution of $2\pi$ is absorbed in the normalization constant $|A|$.
	
\begin{figure}[H]
		\centering
		\includegraphics[scale=0.65]{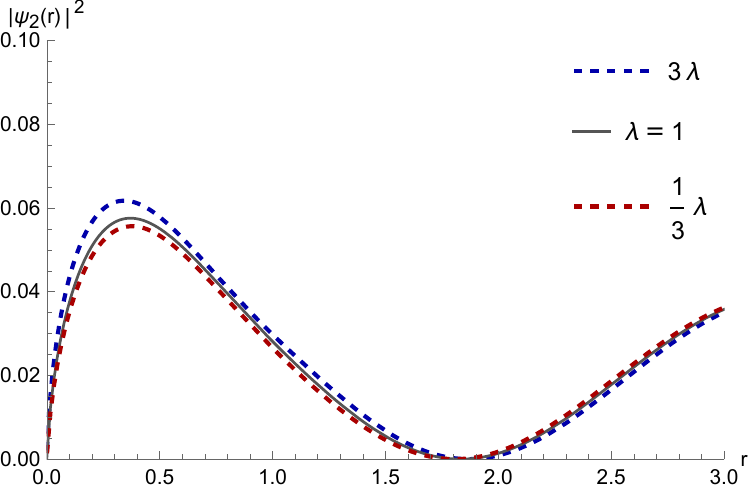}
		\caption{Densities of states for $h_0=b=\mu=1$ and $\epsilon=1$ and $m=\frac{1}{2}$, varying $\lambda$. Where $\psi_2(r)=\zeta(r)\chi_2(r)$. On the blue line, we have an increase in $\lambda$, while on the red line, we have a decrease.}
		\label{fig25}
	\end{figure}

\begin{figure}[H]
		\centering
		\includegraphics[scale=0.65]{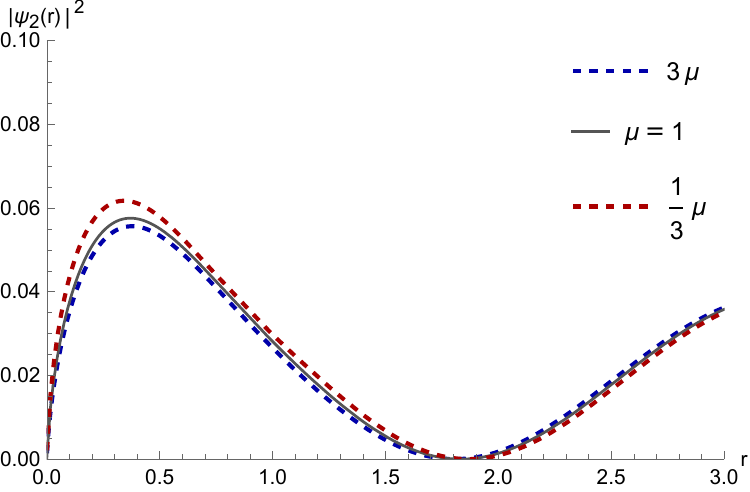}
		\caption{Densities of states for $h_0=b=\lambda=1$ and $\epsilon=1$ and $m=\frac{1}{2}$, varying $\mu$. Where $\psi_2(r)=\zeta(r)\chi_2(r)$. On the blue line, we have an increase in $\mu$, while on the red line, we have a decrease.}
		\label{fig25-2}
	\end{figure}

\begin{figure}[H]
		\centering
		\includegraphics[scale=0.82]{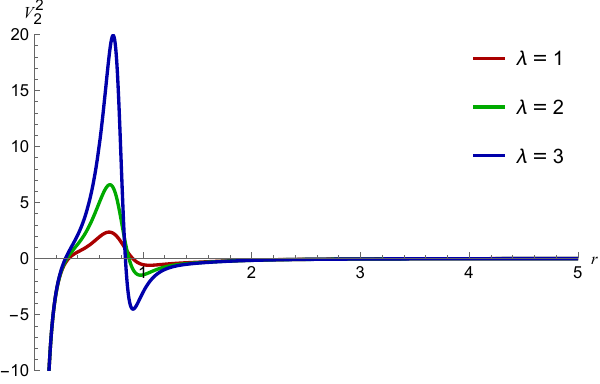}
		\caption{Effective squared potential with $m=\frac{1}{2}$ and $h_0=1$, $b=0.75$, fixed $\mu=1$ and varying $\lambda$. It is observed that the increase in $\lambda$ intensifies the barriers.}
		\label{fig05}
	\end{figure}

\begin{figure}[H]
		\centering
		\includegraphics[scale=0.6]{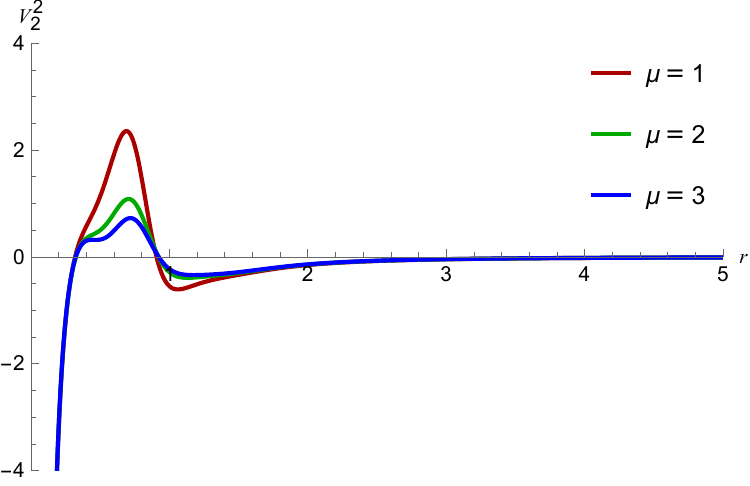}
		\caption{Effective potential with $m=\frac{1}{2}$ and $h_0=1$, $b=0.75$, fixed $\lambda=1$ and varying $\mu$. It is observed that the increase in $\mu$ decreases barriers.}
		\label{fig06}
	\end{figure}

\begin{figure}[H]
		\centering
		\includegraphics[scale=0.65]{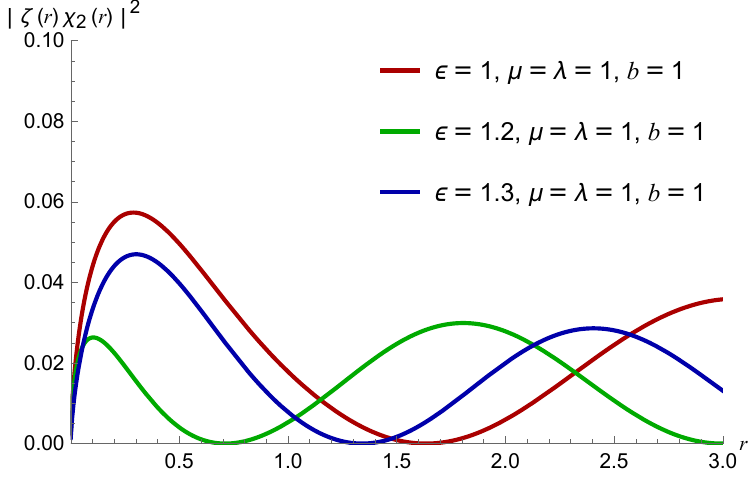}
		\caption{Some solutions with $m=\frac{1}{2}$ and $\mu=\lambda=1$, varying $\epsilon$. The densities undergo more changes near the bump.}
		\label{fig35}
	\end{figure}

	The operator \eqref{20} becomes $\zeta(r)\approx \zeta_0\lim_{h_0\to0}e^{-\int \Gamma_\theta dr}$, so \eqref{30} implies $\psi_2(r)\approx \chi_2(r)$. Due to the complexity of the squared effective potential, we numerically solved equation \eqref{23}. The corresponding probability densities are shown in Figure \ref{fig25}, where the influence of the Lamé coefficients can be seen; we can see that they tend to change the amplitude of the wave functions. Note that the squared effective potential, in figures \ref{fig05} and \ref{fig06}, is not finite and we attribute this to the singularity of the coordinate system at $r=0$, however, it vanishes asymptotically. Observing the figures \ref{fig25} to \ref{fig35}, the elastic parameters would translate into changes in the positions and intensities of the peaks in the local density of states. More clearly, the increase in $\lambda$ is associated with resistance to compressibility in the plane relative to stretching \cite{landau,soutas2012elasticity}, which tends to deepen the attractive region of the effective potential near the deformation. This tends to manifest a higher probability density near the origin. On the other hand, the increase in $\mu$ tends to counterbalance this effect.
	
	It should be mentioned that the explicit contributions of the Lamé coefficients appear from the second-order $\alpha$ approximations. In particular, the solutions are still of the Bessel function type.

	\section{External magnetic field}\label{sect5}
	
	After discussing the strain effects on the electronic states, let us now include an external magnetic field. Assuming an uniform $\vec{B}$ field along the $z$ axis, the vector potential $\vec{A}$ is given by $\vec{\mathcal{A_\theta}}=\frac{1}{2} \vec{B}\times\vec{\mathfrak{r}}$. In cylindrical coordinates, $\vec{\mathfrak{r}}=r\hat{e}_1+z\hat{e}_3$ and $\vec{B}=B_0\hat{e_3}$ yields
	\begin{equation}
		\vec{\mathcal{A}}=\frac{B_0}{2}r\hat{e}_ 2\,.
	\end{equation}
	
\begin{figure}[H]
		\centering
		\includegraphics[scale=0.8]{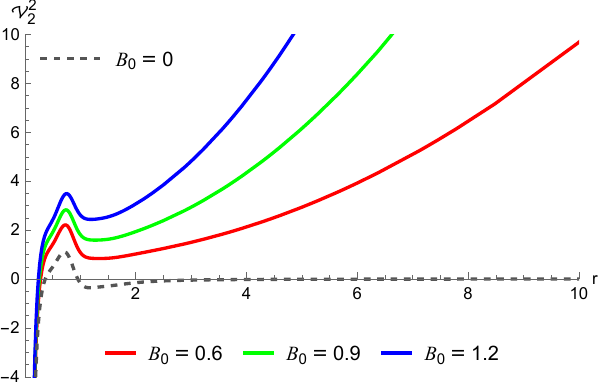}
		\caption{Potentials $\mathcal{V}^2_2$ for some values of $B_0$, in arbitrary units. With $m=\frac{1}{2}$, $h_0=1$, $b=0.8$ and $\mu=\lambda=1$. It is observed that intensity of $B$ alters the asymptotic behavior of the potentials.}
		\label{fig22}
	\end{figure}

\begin{figure}[H]
		\centering
		\includegraphics[scale=0.65]{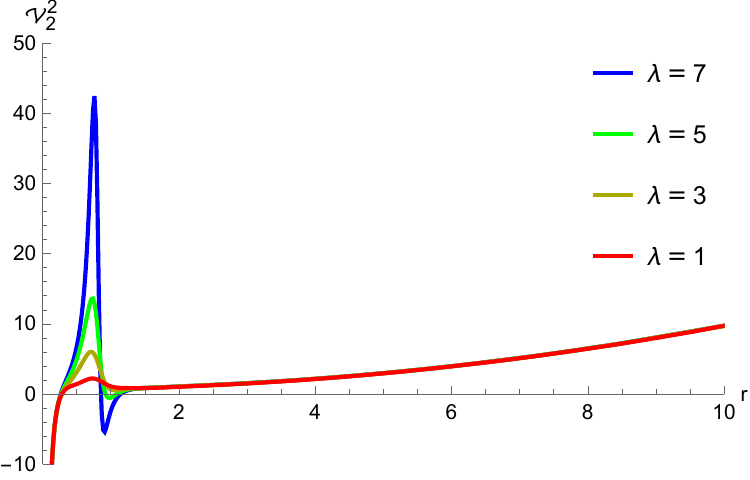}
		\caption{Varying $\lambda$ with $B_0=0.6$, $h_0=1$, $b=0.8$ and $m=\frac{1}{2}$. As in cases without an external field, in figure \ref{fig05}, increasing $\lambda$ increases the potential barrier.}
		\label{fig24}
	\end{figure}

\begin{figure}[H]
		\centering
		\includegraphics[scale=0.65]{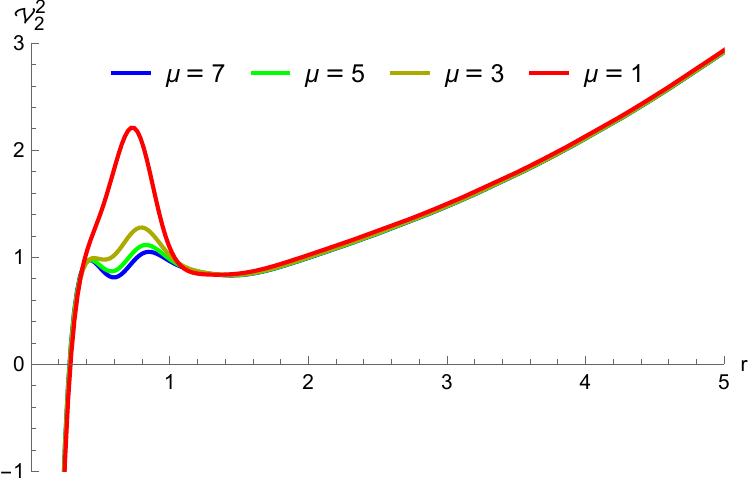}
		\caption{Varying $\mu$ with $B_0=0.6$, $h_0=1$, $b=0.8$ and $m=\frac{1}{2}$. And as discussed in figure \ref{fig06}, the increase in $\mu$ continues to attenuate potential barriers.}
		\label{fig28}
	\end{figure}

	The Hamiltonian for the electron under the influence of strain and magnetic field reads 
	\begin{equation}
		\mathcal{H}=-i\hbar v_F\sigma_3[\gamma^{r}\partial_r+\gamma^\theta(\partial_\theta+\Omega_\theta)+i\gamma^ie\mathcal{A}_i]\,.
	\end{equation}
	Following the same development as in section \ref{sec3}, we will obtain the following equation for $\psi_2$
	\begin{equation}
		-\partial^2_r\psi_2+\left(-2\Gamma_\theta\sqrt{g_{rr}}+\frac{1}{2}\frac{\partial_rg_{rr}}{g_{rr}}\right)\partial_r\psi_2+F^2_2\psi_2=\varepsilon^2\psi_2\,,
	\end{equation}
	with
	\begin{equation}
		\scalebox{1}{$\displaystyle \begin{aligned}
				F^2_2=\frac{g_{rr}m^2}{g_{\theta\theta}}-\frac{m\sqrt{g_{rr}}\partial_rg_{\theta\theta}}{2g^{3/2}_{\theta\theta}}-\sqrt{g_{rr}}\partial_r\Gamma_\theta-g_{rr}\Gamma_\theta^2\\
				-e\frac{2m}{\sqrt{g_{\theta\theta}}}g_{rr}\mathcal{A_{\theta}}-e\sqrt{g_{rr}}\partial_r\mathcal{A_{\theta}}+g_{rr}e^2\mathcal{A}^2_\theta\,.
			\end{aligned}$}
	\end{equation} 
	Where it is clear that the only change concerning \eqref{32} is the inclusion of terms due to the external field in the potential $U^2_2$. That is
	\begin{equation}
		F^2_2=U^2_2+Y^2_2\,,
	\end{equation}
	where
	\begin{equation}
		Y^2_2=-e\frac{2m}{\sqrt{g_{\theta\theta}}}g_{rr}\mathcal{A_{\theta}}-e\sqrt{g_{rr}}\partial_r\mathcal{A_{\theta}}+g_{rr}e^2\mathcal{A}^2_\theta\,.
	\end{equation}
	So the Klein-Gordon equation \eqref{23} in the presence of an external field is
	\begin{equation}\label{31}
		-\chi''_2+\mathcal{V}^2_2\chi_2=\varepsilon^2\chi_2\,,
	\end{equation}
	with $\mathcal{V}^2_2=V^2_2+Y^2_2$. So it is immediately clear that, unlike $\Gamma_\theta$, $\mathcal{A_\theta}$ does not change the holonomy operator \eqref{20}. This is consistent with their distinct physical origins. We can determine a potential vector $\vec{\mathcal{A}}=\frac{B_0}{2}r\hat{e}_ 2$. And by determining the eigenfunctions of \eqref{31}, we can obtain the Landau levels.
	As in \cite{euclides1}, we see that the presence of the external field affects the asymptotic limit of the potentials, while preserving the disturbances due to curvature near the origin. In particular, it is clear that when we turn off the external field, the potentials return to those obtained previously, as can be seen in Figure \ref{fig22}. Figures \ref{fig24} and \ref{fig28} shows how $\lambda$ and $\mu$, that mediated the variation in curvature, alter the interaction with $\vec{B}$, modifying deviations from standard Landau level spacing. And just as in the case without an external field, they continue to behave oppositely, while $\lambda$ tends to increase the barriers, $\mu$ behaves in the opposite manner. These changes can be investigated, for example, via scanning tunnelling microscopy (STM) \cite{ishigami2007,dejuan2011v2}. In particular, the reference \cite{dejuan2011v2} presents experimental proposals for investigating Aharonov-Bohm interference due to fictitious fields originating from strain. In this reference, we can see that these signatures may be oscillations in the local density of states measured by STM and by the dependence on interference with the pseudo-magnetic flux, and therefore with curvature. STM analysis can also accurately estimate the height and width of the bump, as well as the displacement vector.

		\begin{figure}[H]
		\centering
		\includegraphics[scale=0.65]{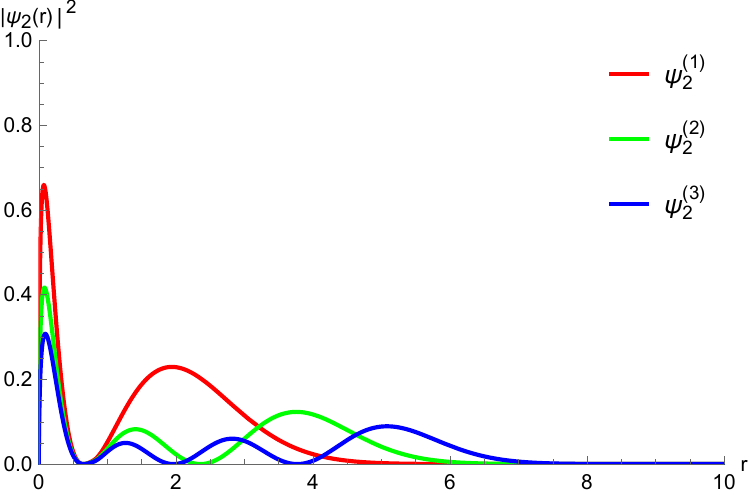}
		\caption{Three first Landau Levels for $B_0=0.6$ in figure \ref{fig22}. Where $b=0.8$, $\lambda=7$, $\mu=1$ and $m=\frac{1}{2}$. It can be observed that as levels increase, the density of states decays more rapidly.}
		\label{correuc03}
	\end{figure}   
	
	\section{Conclusions and Perspectives}
	In this work, we analyzed the effects of a localized Gaussian deformation on massless Dirac fermions confined to a curved surface, incorporating out-of-plane displacements via elasticity theory, and taking into account the explicit contributions of in-plane displacements associated with these transverse modes. We were able to observe that the introduction of the $u_r$ vector proposed by Arias and collaborators \cite{arias1} leads to the appearance of angular contributions in the strain tensor, such that the surface metric is not only changed in the radial component as in \cite{vozmediano1}. Non-linear contributions to $r$ appear in both the spatial components $g_{rr}$ and $g_{\theta\theta}$, revealing nontrivial contributions to the spin connection and curvature. These strain-induced modifications were shown to affect the fermionic density of states, with the Lam\' {e} coefficients playing a significant role in modulating the effective potential. Analytical and numerical results confirmed the emergence of localized states around the deformation, governed by a geometric potential linked to the spinor holonomy. More specifically, since the effective potential is obtained by corrections to the parallel transport of the spinor $\psi_{1,2}$, every geometric contribution can be absorbed in a holonomy operator, as can be seen in \eqref{20} which acts as a geometrical phase on a new function $\chi_{1,2}$ in a similar way to an Aharonov-Bohm phase. Experimentally, this phase could manifest itself in interference experiments, such as ring-shaped deformed graphene geometries or multiple connections, where electrons travel along paths that surround the deformation \cite{dejuan2011,schelter2012,silva2015}. The geometric phase, having a geometric origin, should therefore produce measurable changes in interference fringes or conductance oscillations, even in the absence of a real magnetic flux. Future work may explore scattering processes and transport phenomena associated with these geometric potentials.	
	 We have also added an external magnetic field, and, just as in \cite{euclides1}, it was possible to observe that the explained introduction of $u_r$ does not alter the fact that the Landau levels also shift towards the perturbation.  The effective potential exhibits confining behavior at large distances, leading to the formation of Landau levels that localize near the bump. It was observed that, as levels increase, the decrease in densities intensifies, while behavior near the peak is modulated by geometry. It is important to note that when we have only localized curvature, it generates asymptotically free states, while bound states only arise when we add an external magnetic field, so that the role of curvature focuses on the distribution of states, mediated by the configurations of mechanical parameters.
	
In essence, we obtain that the explicit use of the $u_{r}$ vectors induces the appearance of the Lamé constants $\lambda$ and $\mu$, which are associated with the structure of the lattices intrinsically, as discussed in \cite{katsnelson}, as opposed to the purely geometric perturbations $h(r)$, we also show that using these two approaches it is possible to obtain approximations for the geometric deformation constants in terms of the Lamé constants. We conclude that the Lam\' {e} constants produce changes in the curvature, and therefore in the effective field, and consequently alter the densities of states of the fermions confined to the surface. We obtained analytical and numerical solutions. In both cases, we observed a shift in the densities of states towards the localized perturbation, as previously observed in other geometric configurations \cite{euclides1,yecsiltacs2018dirac,watanabe2015electronic,bueno2012landau,diracplanar}. In particular, asymptotically, we see a behavior mapped by Bessel functions, which is expected due to the cylindrical symmetry of the Gaussian and also from the fact that the geometry is asymptotically flat. By the squared effective potential \eqref{33} we can see that the metric terms are coupled with the angular momentum terms, which indicates a spin-strain coupling.
	For future works, it would also be interesting to study the scattering associated with the potentials analyzed in this work. In particular, as seen in section (\ref{sect4}), a good way of evaluating the scattering cross section would be to apply the Born-Oppenheimer approximation as in \cite{watanabe2015electronic}.

	\section{Acknowledgments}
	Samuel B. B. Almeida thanks to Coordenação de Aperfeiçoamento de Pessoal de Nível Superior (CAPES). J. E. G. Silva thanks the Conselho Nacional de Desenvolvimento Cient\'{i}fico e Tecnolo\'{o}gico (CNPq), grant nº 304120/2021-9. C. A. S. Almeida is supported by grant No. 309553/2021-0 (CNPq/PQ) and by Project UNI-00210-00230.01.00/23 (FUNCAP).
	
	\section*{CONFLICTS OF INTEREST/COMPETING INTEREST}

The authors declared that there is no conflict of interest in this manuscript. 

\section*{DATA AVAILABILITY}

No data was used for the research described in this article.	
	
	
	\appendix
		\section{Determining geometric parameters in mechanical terms}\label{relationworks}
\begin{figure}[H]
		\centering
		\includegraphics[scale=0.58]{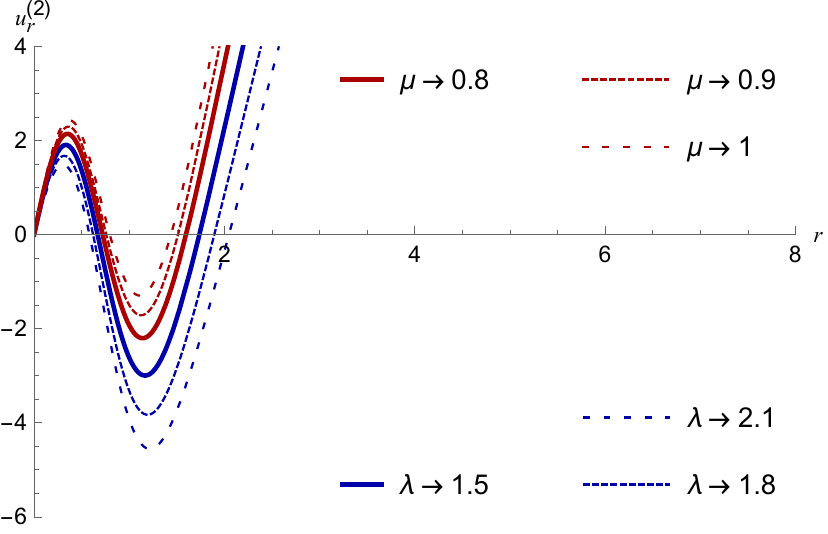}
		\caption{Changes in $\lambda$ and $\mu$ in $u^{\text{(2)}}_r$. The approximation is not asymptotically zero as expected from a localized deformation. However, close to the origin, it behaves similarly to the exact result in figure \ref{fig07}.}
		\label{variaçãounovo}
	\end{figure}

	In the work \cite{vozmediano1}, Fernando and colleagues used the purely geometric method to describe Gaussian perturbations on a graphene surface through
	\begin{equation}\label{metricvozmediano}
		ds^2= \left(1+\alpha^2f(r)\right) dr^2+r^2d\theta^2\,, \alpha=\frac{h_0}{b}\,.
	\end{equation}
	In this way, the lamé constants that characterize the mechanical properties of the material do not appear explicitly. Using the displacement suggested in \cite{arias1}, we were able to obtain an approximation for the constant $\alpha$ in terms of the Lamé coefficients. To do this, we expand the displacement \eqref{04} in first order, which gives us
	\begin{equation}\label{expandArias}
		u_r(r)\approx\frac{\sqrt{\frac{\pi }{2}} h_0^2 \mu  }{b^2 (\lambda +2 \mu )}r+\mathcal{O}(r^1,r^3,r^5...)\,,
	\end{equation}
	it should be mentioned that reference \cite{arias1} obtains this displacement by minimizing the elastic Hamiltonian associated with the surface. However, by linearizing the equations, we can consider another way to obtain the displacements associated with the transverse Gaussian perturbation and develop an approximation. 
	In particular, we can start from the stress tensor, which can be written as
	\begin{equation}
		\sigma_{\mu\nu}=\lambda u_\kappa^{\ \kappa}g_{\mu\nu}+2\mu u_{\mu\nu}\,,\text{ where $u_\rho^{\ \rho}=g^{\gamma\beta}u_{\gamma\beta}$.}
	\end{equation}
	In first order we have
	\begin{equation}
		g^{\gamma\beta}=(\delta_{\gamma\beta}+2u_{\gamma\beta})^{-1}\approx \delta^{\gamma\beta}-2u^{\gamma\beta}+\mathcal{O}(u^2)\,,
	\end{equation}
	so in this regime
	\begin{equation}
		\sigma_{\mu\nu}=\lambda u_\kappa^{\ \kappa}\delta_{\mu\nu}+2\mu u_{\mu\nu}\,.
	\end{equation}
	Let's assume that there are no external forces acting on the surface, in which case the equilibrium equation is
	\begin{equation}
		\nabla_\nu\sigma_{\mu\nu}=0 \quad \text{, with $\mu=r$, $\theta$}\,.
	\end{equation}
	Developing the covariant derivative, and considering the regime where the $\mathcal{O}(u^2)$ contributions are neglected, the equilibrium equation reduces to
	\begin{equation}
		\partial^2_ru_r+\partial_r(\partial_rh)^2=0\,,
	\end{equation}
	for the transversal perturbation in \eqref{01} the solutions of this equation are
	\begin{equation}
		u^{\text{(2)}}_r(r)=C_4+C_5 r-\frac{h_0^2 \left(\sqrt{2 \pi } b \operatorname{Erf}\left(\frac{\sqrt{2} r}{b}\right)-4 r e^{-\frac{2 r^2}{b^2}}\right)}{4 b^2}\,.
	\end{equation}
	Looking at \eqref{04}, we see that $u_r$ converges to zero, so by consistency, let's take $C_4=0$, so
	\begin{equation}
		u^{\text{(2)}}_r(r)=C_5 r-\frac{h_0^2 \left(\sqrt{2 \pi } b \operatorname{Erf}\left(\frac{\sqrt{2} r}{b}\right)-4 r e^{-\frac{2 r^2}{b^2}}\right)}{4 b^2}\,.
	\end{equation}
	The behavior of this approximation is similar to that of \eqref{04} near the origin, as illustrated in Figures \ref{fig07} and \ref{variaçãounovo}. However, it is more sensitive to changes in $\lambda$ and $\mu$. 
	
	As mentioned in section \ref{sect2}, increases in $\mu$ tend to elevate contributions $	u^{\text{(2)}}_r>0$, while $\lambda$ behaves oppositely, which is the same behavior as $u_r$. Moreover $u^{\text{(2)}}_r$ does not asymptotically approach zero. We can rename $C_5=\beta$, and then calculate the metric with this displacement, we will find
	\begin{equation}
		ds^2=(1+2\beta)dr^2+r^2d\theta^2\,,
	\end{equation}
	i.e., the disturbances are constant in this regime. Also expanding $u^{\text{(2)}}_r(r)$ we have
	\begin{equation}
		u^{\text{(2)}}_r(r)\approx\beta r+\mathcal{O}(r^1,r^3,r^5...)\,,
	\end{equation}
	so comparing with \eqref{expandArias} we conclude that
	\begin{equation}
		\beta=\alpha^2\frac{\sqrt{\frac{\pi }{2}} \mu  }{ (\lambda +2 \mu )}\,,
	\end{equation}
	or alternatively
	\begin{equation}
		\alpha^2=\frac{(\lambda +2 \mu )}{\mu}\beta\,.
	\end{equation}
	So the expression \eqref{metricvozmediano}, used in \cite{vozmediano1}, can be written in terms of the Lamé parameters by
	\begin{equation}
		ds^2= \left[1+\beta\frac{(\lambda +2 \mu )}{\mu} f(r)\right] dr^2+r^2d\theta^2\,.
	\end{equation}
	it is interesting to note the persistent presence of the term $\lambda+2\mu$, in some areas of research, this quantity is defined as uniaxial stress \cite{aki2002quantitative, mavko2020rock}.
Although it correctly captures the behavior near the source and allows an analytical connection to the purely geometric parameter $\alpha$, the displacement $u^{(2)}_r$ is less suitable for global and localization spectral analyses due to this non-zero asymptotic contribution. For this reason, in order to analyze a more general case, in this work we use $u_r$ defined by Eq. \eqref{04} because it reproduces the expected null asymptotic behavior of a localized deformation and avoids long-range non-physical metric distortions.
It is important to emphasize that obtaining the exact value of $u_r$ using the approach described in \cite{arias1} approach is completely non-trivial, since it is necessary to use the F\"{o}ppl-Von-Kármán procedure \cite{landau}, which, by involving the minimization of the elastic energy of the surface, introduces higher-order terms in the deformation-displacement relationship and is capable of fully capturing the stretching effect that occurs when membranes are significantly bent.  In this case, the out-plane displacements are coupled to the in-plane motion through fourth-order nonlinear equations called F\"{o}ppl equations. Therefore, the procedure for linearizing the equilibrium equations discussed here may be useful in future work in order to simplify the equations.

	\begin{widetext}	
	\section{Explicit expressions}\label{apendd}
	By expanding the  metrics \eqref{07} and \eqref{08} in second order in $\alpha$, the Lamé coefficients and geometric parameters of the Gaussian can be absorbed into
	\begin{equation}\label{ccst01}
		C_1=h^2_0\frac{\left(3 \sqrt{2 \pi } \lambda -18 \sqrt{2 \pi } \mu +32 \mu \right)}{4\mu b^4}
	\end{equation}
	\begin{equation}\label{ccst02}
		C_2=-\frac{\sqrt{\frac{\pi }{2}} h_0^2 \lambda }{2 b^2 \mu }+\frac{\sqrt{\pi } h_0^2}{\sqrt{2} b^2}
	\end{equation}
	\begin{equation}\label{ccst03}
		C_3=1+\frac{\sqrt{2 \pi } h_0^2 \mu }{b^2 (\lambda +2 \mu )}
	\end{equation}
	The corrections to the parallel transport of the spinor are introduced into the spin connection via the following Christoffel symbols
	\begin{equation}\label{crsft01}
		\scalebox{0.9}{$\displaystyle \begin{aligned}
		\Gamma^\theta_{r\theta}=\frac{2 r \left(\sqrt{2 \pi } h_0^2 \left(b^2 \mu -2 r^2 (\lambda +2 \mu )\right)+b^4 e^{\frac{2 r^2}{b^2}} (\lambda +2 \mu )\right)}{\sqrt{2 \pi } h_0^2 b^2 \left(2 r^2 (\lambda +2 \mu )-b^2 \left(e^{\frac{2 r^2}{b^2}}-1\right) (\lambda +\mu )\right)+2 b^4 r^2 e^{\frac{2 r^2}{b^2}} (\lambda +2 \mu )}
			\end{aligned}$}
	\end{equation}
	\begin{equation}\label{crsft02}
		\scalebox{0.9}{$\displaystyle\begin{aligned}
				\Gamma^r_{\theta\theta}=\frac{2 r^3 \left(\sqrt{2 \pi } h_0^2 \left(b^2 \mu -2 r^2 (\lambda +2 \mu )\right)+b^4 e^{\frac{2 r^2}{b^2}} (\lambda +2 \mu )\right)}{h_0^2 \left(2 \sqrt{2 \pi } b^2 \lambda  r^2-\left(\sqrt{2 \pi } b^4 \left(e^{\frac{2 r^2}{b^2}}-1\right) (\lambda +\mu )\right)+8 \left(\sqrt{2 \pi }-2\right) r^4 (\lambda +2 \mu )\right)-2 b^4 r^2 e^{\frac{2 r^2}{b^2}} (\lambda +2 \mu )}
			\end{aligned}$}
	\end{equation}
	\begin{equation}\label{crsft03}
		\scalebox{0.9}{$\begin{aligned}
				\Gamma^r_{rr}=-\frac{2 h_0^2 \left(-2 \sqrt{2 \pi } b^4 r^2 (\lambda +\mu )+4 b^2 r^4 \left(\left(\sqrt{2 \pi }-4\right) \lambda +4 \left(\sqrt{2 \pi }-2\right) \mu \right)\right)}{h_0^2 b^2 r \left(-2 \sqrt{2 \pi } b^2 \lambda  r^2+\sqrt{2 \pi } b^4 \left(e^{\frac{2 r^2}{b^2}}-1\right) (\lambda +\mu )-8 \left(\sqrt{2 \pi }-2\right) r^4 (\lambda +2 \mu )\right)+2 b^6 r^3 e^{\frac{2 r^2}{b^2}} (\lambda +2 \mu )}+\\
				+\frac{\sqrt{2 \pi } b^6 \left(e^{\frac{2 r^2}{b^2}}-1\right) (\lambda +\mu )-16 \left(\sqrt{2 \pi }-2\right) r^6 (\lambda +2 \mu )}{h_0^2 b^2 r \left(-2 \sqrt{2 \pi } b^2 \lambda  r^2+\sqrt{2 \pi } b^4 \left(e^{\frac{2 r^2}{b^2}}-1\right) (\lambda +\mu )-8 \left(\sqrt{2 \pi }-2\right) r^4 (\lambda +2 \mu )\right)+2 b^6 r^3 e^{\frac{2 r^2}{b^2}} (\lambda +2 \mu )}
			\end{aligned}$}
	\end{equation}
	We can immediately see the algebraic complexity that naturally requires numerical treatment. In particular, it is more insightful to write the disturbances in terms of the displacement vector $u_r$, as we can see in the following curvature case
	
	\begin{equation}\label{24}
		\scalebox{0.9}{$
			\begin{aligned}
				\mathcal{R}=-\frac{1}{r^2 (r^3 + 2 u_r)^2 (1 + 2 (\partial_r h)^2 + 2 (\partial_r u_r))^2} \Bigg[
				2 (1 + 2 (\partial_r h)^2 + 2 (\partial_r u_r)) \Bigg(-3 u_r (2 r^3 + u_r) + 2 r (2 r^3 + u_r) \partial_r u_r + r^2 (\partial_r u_r)^2\Bigg)\\
				+ 4 r (r^3 + 2 u_r) \partial_r h \Bigg(r^3 - u_r + r \partial_r u_r\Bigg) \partial_r^2 h
				+ 2 r (r^3 + 2 u_r)\Bigg(-u_r + r \Bigg(-1 + r^2 - 2 (\partial_r h)^2 - \partial_r u_r\Bigg)\Bigg) \partial_r^2 u_r\Bigg]\,.
			\end{aligned}
			$}
	\end{equation}
	\end{widetext}

\end{document}